\documentclass[conference]{IEEEtran}
\usepackage{amsmath,amssymb,amsfonts}
\usepackage{algorithmic}
\usepackage{graphicx}
\usepackage{textcomp}
\usepackage{xcolor}
\usepackage{svg}
\usepackage{float}
\usepackage{tcolorbox}
\usepackage{url}
\usepackage{pgfplots}
\usepackage{pgfplotstable}
\usepackage[hidelinks]{hyperref}
\usepackage{filecontents}
\usepackage{tikz}
\usepackage{float}
\usepackage[
sorting=none,
]{biblatex}

\definecolor{babyblueeyes}{rgb}{0.63, 0.79, 0.95}
\definecolor{chad-green}{RGB}{79,159,98}

\usetikzlibrary{patterns}
\usetikzlibrary{positioning}

\def\BibTeX{{\rm B\kern-.05em{\sc i\kern-.025em b}\kern-.08em
    T\kern-.1667em\lower.7ex\hbox{E}\kern-.125emX}}

\bibliography{references}
\begin{document}

\title{Joint Embeddings for Graph Instruction Tuning}

\author{\IEEEauthorblockN{ Aaron Haag}
\IEEEauthorblockA{\textit{Siemens Technology Department} \\
Munich, Germany \\
aaron.haag@siemens.com}
\and
\IEEEauthorblockN{Vlad Argatu }
\IEEEauthorblockA{\textit{Siemens Technology Department} \\
Munich, Germany \\
vlad.argatu@siemens.com}
\and
\IEEEauthorblockN{ Oliver Lohse}
\IEEEauthorblockA{\textit{Siemens Technology Department} \\
Munich, Germany \\
oliver.lohse@siemens.com}
}

\maketitle

\begin{abstract}
Large Language Models (LLMs) have achieved impressive performance in text understanding and have become an essential tool for building smart assistants.
Originally focusing on text, they have been enhanced with multimodal capabilities in recent works that successfully built visual instruction following assistants.
As far as the graph modality goes, however, no such assistants have yet been developed.
Graph structures are complex in that they represent relation between different features and are permutation invariant.
Moreover, representing them in purely textual form does not always lead to good LLM performance even for finetuned models.
As a result, there is a need to develop a new method to integrate graphs in LLMs for general graph understanding.
This work explores the integration of the graph modality in LLM for general graph instruction following tasks.
It aims at producing a deep learning model that enhances an underlying LLM with graph embeddings and trains it to understand them and to produce, given an instruction, an answer grounded in the graph representation.
The approach performs significantly better than a graph to text approach and remains consistent even for larger graphs.

\end{abstract}


\section{Introduction}

Large Language Models (LLMs) have become an important method for problem solving using artificial intelligence mainly due to their generative capabilities that make them able to effectively follow user instructions.
Hence, they are ideal models for designing and creating chat bots that can effectively interact with an user.
Unfortunately, since LLMs operate on text, most of the chat bots are limited to this modality.
This is the reason why, in recent years, significant effort has been made to enhance LLMs with multimodal input.
It has especially been the case for image \cite{llava} \cite{li2023blip2} \cite{dai2023instructblip} \cite{zhu2023minigpt4} \cite{song2024moma}, video \cite{xu2024pllava} \cite{shu2023audiovisual} and graph \cite{llm-on-graphs}.
Graph structures are especially important within industrial applications, like e.g. implementation of programmable software controllers (PLCs) \cite{pavlovskyi2018template} or Computer-Aided-Design (CAD) representations \cite{COLLIGAN2022103226cadnet}.
Integrating graphs in LLMs, with their structure which is permutation invariant and always represents some kind of relationship, is a highly challenging task.
To achieve this goal, a natural approach is to leverage LLMs understanding of structured input by representing graphs, or subgraphs, directly as text \cite{ye2024language} \cite{llm-vs-graph} \cite{chen2024exploring} \cite{chen2024graphwiz}.
This approach has the advantage of using mainly in-context learning and thus requiring little to no training.
However, it raises the question of the textual representation of a graph.
Other methods rely on using learned embedding representation for node features \cite{graph-gpt} \cite{chen2024llaga} or directly for the entire graph \cite{graph-llm}.

This paper introduces a new method for graph instruction tuning of LLMs, which is the process of finetuning them for the instruction following task, by enhancing them with graph understanding capabilities.
Inspired by the success of \cite{llava} and its ability to scale to modern architectures \cite{liu2023improved} \cite{liu2024llavanext} \cite{lin2024moellava}, the new method converts a graph into a fixed number of embeddings and injects them into a LLM alongside an user instruction.
The LLM is trained to understand the embeddings coming from the graph and to use them to formulate a correct answer to the user's instruction.
This method outperforms the graph to text approach and does not suffer from performance decay when the size of the graph increases.
Moreover, as it only acts at the embedding layer, it is agnostic to the LLM architecture used as backbone and is thus more scalable than \cite{graph-llm}.

The paper is composed of the following sections:
\begin{itemize}
    \item A \textsc{Related Work} section that presents the architectures that inspired the proposed approach and those trying to solve similar problems.
    \item A \textsc{Method} section that introduces the architecture as well as the training procedure.
    \item A \textsc{Numerical Results} section that explains the overall experimental setup and comments the results.
    \item A \textsc{Limitations} section that discusses the limitations of the proposed framework and of the obtained results.
    \item A \textsc{Conclusion} that summarizes the results and suggests future works and improvements.
\end{itemize}

\section{Related Work}

This section first describes the visual integration approach that inspired the graph method before describing similar methods to the proposed architecture.

\subsection{Integration of Vision in Language Models}

Vision language models, which extend LLMs by incorporating images, have performed well in visual question answering tasks \cite{llava} \cite{li2023blip2} \cite{dai2023instructblip} \cite{zhu2023minigpt4} \cite{li2022blip}.
Among them, \cite{llava} proposes a model that accepts two different modalities, an image, and text that is usually an instruction on the inputted image.
To combine those inputs, a LLM is enhanced with an image encoder.

To encode the image, a visual transformer (ViT) \cite{vit} encoder trained in a contrastive Language-Image setup \cite{clip} is leveraged.
Contrastive Language-Image learning aims at aligning the embeddings of a visual encoder and a textual encoder so that the euclidean distance between an image and its textual description are as close as possible.
After training, the visual encoder is thus able to project its input into a space that follows a structure relevant from the textual embedding point of view.
Projection from this space into any other textual embedding space is thus, intuitively, easier.

The multimodal model embeds the textual input with the LLM embedding layer and the image with the ViT encoder.
The image embeddings are then aligned with the LLMs embedding space using a MLP and merged with the textual embeddings.
The multimodal embeddings are passed to the rest of the LLM that is fine-tuned on image instruction answering and is thus learning to use the newly added embeddings to generate an answer to the user's instruction.

This architecture manages to perform very well on zero-shot question answering tasks showing that direct image integration is a viable way of incorporating this data type.
Many direct improvements have been proposed \cite{liu2023improved} \cite{liu2024llavanext}, using better backbone LLMs and more advanced datasets.
Other improvements focus on the model's architecture by either modifying the encoder part, as done in \cite{hu2023bliva} where the encoder part is replaced by an InstructBLIP \cite{dai2023instructblip} architecture, or upgrading the backbone LLM to more modern architectures as done in \cite{lin2024moellava} that uses a mixture of experts \cite{moe} \cite{eigen2014learning} architecture.

The approach introduced in \cite{llava} creates a highly scalable bridge between different modalities and text. It has also been adapted to video \cite{shu2023audiovisual} \cite{xu2024pllava} and image generation \cite{song2024moma}.

\begin{figure}
    \centering
    \includegraphics[scale=0.45]{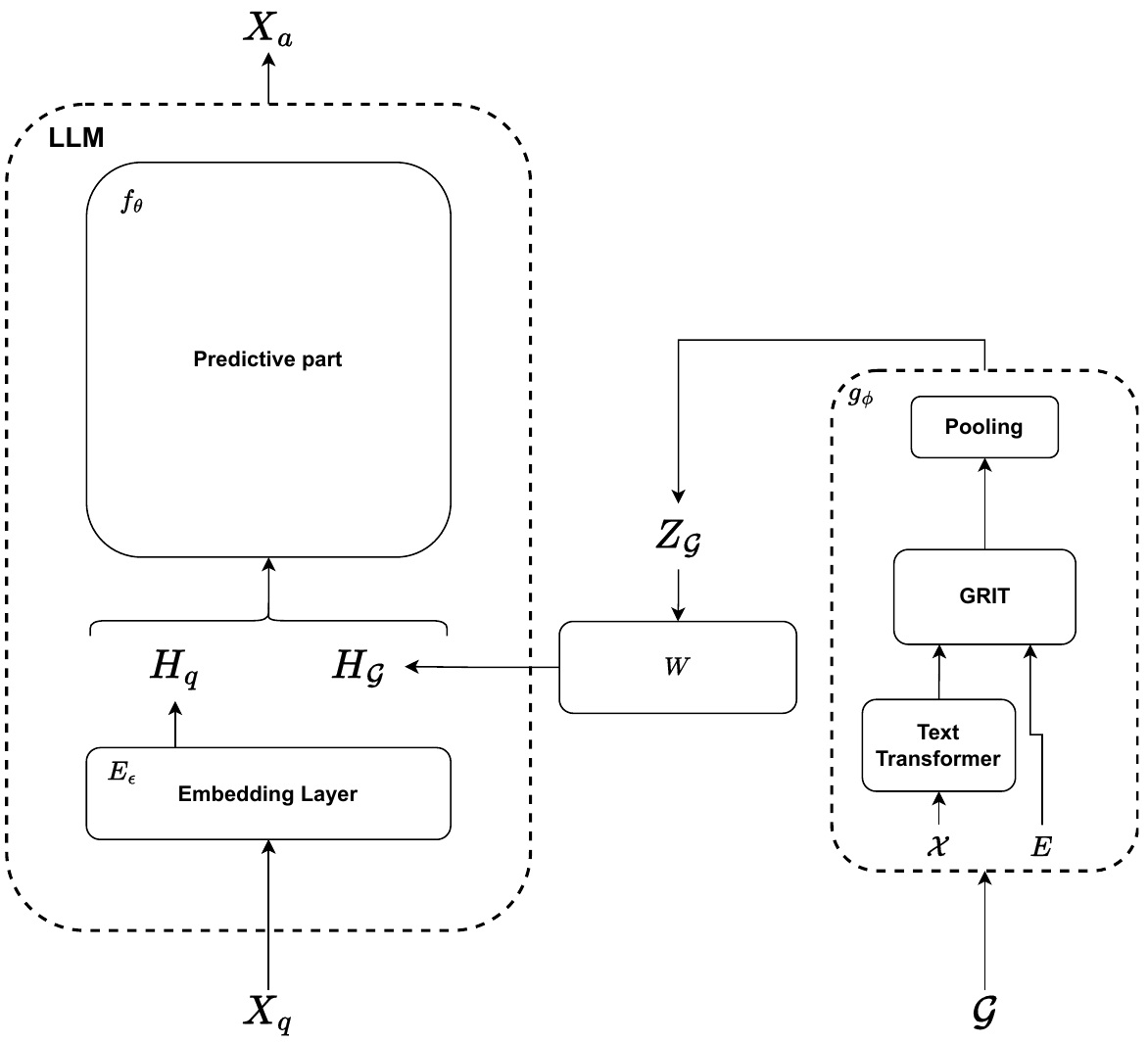}
    \caption{GraphLlava architecture. The Graph encoder part is adapted from \cite{graph-llm}. The used Graph Transformer is introduced in \cite{Ma2023GraphIB} and is known as Graph Inductive bias Transformer (GRIT).}
    \label{gllava_arch}
\end{figure}

\subsection{Integration of Graphs in Language Models}

A common method for integrating graphs in large language model is to represent them directly as text \cite{ye2024language} \cite{llm-vs-graph} \cite{chen2024exploring} \cite{chen2024graphwiz}.
This method has the advantage of using directly in-context learning and is thus computationally efficient.
It poses, on the other hand, the problem of graphical textual representation.
In fact, graphs can have very complex node features going from mathematical vectors to text.
Representing a graph in a textual format would require finding a way to describe the node features as well as the relationships between the nodes.
For graphs having complex node features and a large amount of edges, this approach ends-up in generating large prompts on which most LLMs have limited understanding capabilities \cite{liu2024lostinmiddle}. 

The successful approach developed in \cite{llava} for image integration in LLMs inspired a series of graph integration techniques that use adapters to inject graph embeddings directly in the LLM architecture \cite{graph-gpt} \cite{chen2024llaga} \cite{graph-llm}.
These techniques do not depend on textual representation of the underlying structure and do not require to define a method of representing graph in textual form.

In \cite{graph-gpt} the authors train a node features to text encoder in a contrastive learning setup that is close to the one introduced by CLIP \cite{clip}.
This encoder transforms node features to embeddings that are then converted to LLM features using a MLP. 
For a specific user instruction, a certain number of nodes are inputted and encoded with the node encoder.
The instruction's embeddings and node embeddings are merged and passed to the LLM archictecture.
This approach requires training a contrastive learning encoder and thus having a graph dataset that gives a textual description of each node in addition to relevant node features.
\cite{chen2024llaga} uses a similar approach except it gets rid of the contrastive encoder and directly aligns node embeddings (obtained using a text encoder) with the textual LLM embeddings.

Both approaches have as a downside the fact that they operate at node level and are trained for node classification and link prediction while our work aims at producing a general purpose graph assistant.
Such assistant should also be able to answer questions about the overall graph structure.
Such questions would require the models proposed by \cite{graph-gpt} and \cite{chen2024llaga} to use most of node representations as embeddings for the LLM hence having a similar scaling problem as graph to text approaches.

The approach in \cite{graph-llm} proposes a LLM adapter for graphö features.
It does so by training a graph encoder that produces, for each attention layer of a backbone transformer based LLM, a sequence of prefix embeddings that are added to the queries and keys.
This architecture operates on graphs using textual features as node features and trains only the adapter.
Although very promising, the model proposed in \cite{graph-llm} is not trained on a diverse dataset of questions and thus does not have relevant results for the general graph question answering task.
In fact, one different model is trained for each different question.
Moreover, its adapter architecture requires the modification of the attention computation mechanism and is, as a consequence, less resilient to changes in the backbone LLM architecture.
It is only usable with transformer architectures using attention modules as defined in \cite{vaswani2023attention} and thus can not be used with newer approaches such as selective state spaces \cite{gu2023mamba}.

\section{Method}

\subsection{GraphLlava architecture}

Inspired by the approach introduced in \cite{llava}, we enhance a pretrained LLM  with graph embeddings produced by a pretrained graph encoder.
The architecture is described in Fig. \ref{gllava_arch}.
We name this architecture GraphLlava.

Formally, the architecture is composed of a LLM that is caracterized by its predictive part $f_{\theta}$ and by the textual embedding layer $E_{\epsilon}$.
The LLM is enhanced with a graph encoder $g_{\phi}$ that, given a graph, produces graph embeddings.
The graph embeddings are projected in the textual embedding space using the projection matrix $W$.

The overall goal is to train the language model and the projection matrix to answer a user query given a graph.
Let $\mathcal{G} = (\mathcal{X}, E)$ be the inputted graph, $X_q$ the user's query and $X_a$ the ground truth answer.
The set $\mathcal{X}$ depicts the textual node features of the graph and $E$ represents the edges.
The final model has to generate $X_a$ given $X_q$ and $\mathcal{G}$.

To this end, the first step is to compute a graph representation that can be used by the LLM.
Graph embeddings are thus computed using $g_{\phi}$.
Let $Z_{\mathcal{G}}$ be the graph embeddings of $\mathcal{G}$, they are computed using equation (\ref{eq:graph_embed}).

\begin{equation}
   Z_{\mathcal{G}} = g_{\phi}(\mathcal{G})
   \label{eq:graph_embed}
\end{equation}

The graph embeddings are projected into the textual embedding space using the $W$ projection matrix to produce $H_{\mathcal{G}}$, the textual space representation of $\mathcal{G}$.
This operation is described in equation (\ref{eq:ztoh}).
\begin{equation}
    H_{\mathcal{G}} = W \cdot Z_{\mathcal{G}}
    \label{eq:ztoh}
\end{equation}

The user query $X_q$ is embedded using the embedding layers to produce $H_q$, the textual embeddings of $X_q$.
This operation is described in equation (\ref{eq:qembed}).
The representation of the user query $H_q$ is concatenated with $H_{\mathcal{G}}$ to form a new sequence of embeddings that represents both user's query $X_q$ and the inputted graph $\mathcal{G}$.
\begin{equation}
    H_q = E_{\epsilon}(X_q)
    \label{eq:qembed}
\end{equation}

\begin{figure}
\centering

\begin{tcolorbox}

\begin{itemize}[]

\item ``Describe the graph concisely."
\item ``Provide a brief description of the given graph."
\item ``Offer a succinct explanation of the graph presented."
\item ``Summarize the content of the graph."
\item ``Give a short and clear explanation of the subsequent graph."
\item ``Present a compact description of the graph key features."
\item ``Render a clear and concise summary of the graph."
\item ``Write a terse but informative summary of the graph."
\item ``Create a compact narrative representing the graph presented."
    
\end{itemize}

\end{tcolorbox}
    
\caption{The list of instructions for brief graph description. This list is the same as the one proposed for image instruction tuning in \cite{llava} but adapted to ask for graph description.}
    \label{concise_description}

\end{figure}

\begin{figure}
\centering

\begin{tcolorbox} 

\textbf{SYSTEM}: $X_{system-message}$  EOS\_TOKEN \\
\textbf{USER}: $X_q$ EOS\_TOKEN \\
\textbf{ASSISTANT}: \textcolor{magenta}{$X_a$ EOS\_TOKEN}

\end{tcolorbox}
    
\caption{Training input format. Loss is computed only on the colored tokens.}
    \label{tiformat}

\end{figure}

The concatenated embeddings are given to the predictive part of the LLM, $f_{\theta}$, that will generate $X_a$, the answer to the user query.
This computation and the overall process are described in equation (\ref{eq:allpipeline}).

\begin{equation}
    \begin{aligned}
    X_a & = f_{\theta}(H_q \| H_g)\\
        & = f_{\theta}(E_{\epsilon}(X_q) \| W \cdot g_{\phi}(\mathcal{G}))
    \end{aligned}
    \label{eq:allpipeline}
\end{equation}

\subsection{Extracting Graph Embeddings}

To extract the graph embeddings, a graph encoder pretrained on a downstream task that involves the use of graph textual representation is required.
An encoder similar to the one introduced in \cite{graph-llm} is pretrained on a graph question answering task and is leveraged as $g_{\phi}$.

The training framework described in \cite{graph-llm} is modified so that the final model is not specific to the inputted question.
The encoder is trained on a graph question answering dataset where each question asks to describe the graph and the ground truth is the textual description of the graph.
In order to have a diverse set of input questions, for each graph, a question is uniformly sampled from the set described in Fig. \ref{concise_description}.
This set is similar to the one introduced in \cite{llava} for visual question answering but adapted for graphs.

After training, the encoder is slightly modified to fit the needs of the newly introduced architecture.
In \cite{graph-llm}, the graph encoder produces an embedding for each attention layer of the transformer architecture of the underlying LLM.
This approach leads to an overall architecture that is more difficult to scale to newer LLM architectures as it requires modifying the logic behind the attention computation.
That is why a mean pooling layer is added so that the encoder produces a single embedding tensor that is used as $Z_{\mathcal{G}}$.

\subsection{Training procedure}

A two stage training routine is used.
In both training stages, the training instructions follow the format described in Fig. \ref{tiformat}.

\textbf{Stage 1: Pre-Training for Feature Alignement.} This part aims at aligning $W$ so that the conversion from $Z_{\mathcal{G}}$ to $H_{\mathcal{G}}$ produces embeddings that can be used by the LLM.
To achieve the desired goal, all network's parameters except $W$ are frozen.
The trainable parameters are thus $\psi = \{W\}$.
The projection network is trained on a dataset of tuples $\left( \mathcal{G}, X_q, X_a\right)$ with $X_q$ being an instruction asking to describe the given graph. This instruction is uniformly sampled from the set described in Fig. \ref{concise_description}. 
The ground truth, $X_a$, is a textual description of the graph $\mathcal{G}$.

\textbf{Stage 2: Fine-Tuning End-to-End.} We freeze $g_{\phi}$ and train the LLM and the projection matrix. i.e the trainable parameters are $\psi = \{ \theta, \epsilon, W\}$. The model is trained on single turn conversations. For this stage, the dataset is made of tuples $\left(  \mathcal{G}, X_q,  X_a\right)$ with $X_q$ being a question about the graph $\mathcal{G}$ and $X_a$ its answer.

Using the standard LLM autoregressive modelling, the process can be described by equation (\ref{eq:autoregressive}).
\begin{equation}
    p(X_a | \mathcal{G}, X_q) = \prod_{i = 0}^L p_{\psi}(x_i |\mathcal{G}, X_{q}, X_{a, <i})
    \label{eq:autoregressive}
\end{equation}

Where $L$ is the length of the sequence, $\psi$ is the set of the trainable parameters, $X_{a, <i}$ are respectively the answer tokens before the predicted token $x_i$.
The training objective aims at maximizing this probability for the tokens that need to be generated.
The $\mathcal{G}$ is added to the standard autoregressive equation to put emphasis on the fact that each token is generated having access to the entire graph representation.
In other words, the answer is grounded on the graph representation.

\section{Numerical Results}
\subsection{Used Dataset and implementation}

A dataset $\mathcal{D}$ is defined in equation (\ref{eq:ds}).
\begin{equation}
    \mathcal{D} = \{\left( \mathcal{G}_i, X_q^i, X_a^i\right), i \in \left[ 0, N\right]  \}
\hspace{2px} \textrm{with} \hspace{2px} \forall i, \mathcal{G}_i = (\mathcal{X}_i, E_i)
    \label{eq:ds}
\end{equation}
Where $N$ is the number of elements in the dataset, $\mathcal{G}_i$ a graph, $\mathcal{X}_i$ are the node features of $\mathcal{G}_i$ represented as textual tokens, $E_i$ are the edges of $\mathcal{G}_i$, $X_q^i$ is an instruction and $X_a^i$ is the expected answer.
In practice, each node content is tokenized with the same tokenizer used to tokenize $X_q^i$ and $X_a^i$.
Due to limited GPU memory available at training, $X_a^i$ and $X_q^i$ are cropped so that concatenating a $X_q^i$ and a $X_a^i$ results in a sequence of at most 512 tokens. 

\begin{figure}
\centering

\begin{tcolorbox} 

\textbf{SYSTEM}: ``You are an expert on graphs. You will have a query that is a question on graphs, a ground truth that is the expected answer and two answers coming from user1 and user2. You will have to choose which one is as good as the ground truth. Please answer only with the words 'user1' or 'user2' or 'none'"\\
\textbf{USER}: ``The query is \{$X_q$\}. The ground truth is: \{$X_a$\}. user1's answer is \{$Y_{gllava}$\}. user2's answer is \{$Y_{tllama}$\}."

\end{tcolorbox}
    
\caption{Prompt format for GPT-4-as-judge prompting. $X_q$, $X_a$, $Y_{gllava}$ and $Y_{tllama}$ being respectively the input query containing the textual description of the graph, the ground truth answer, the answer given by the GraphLlava model and the answer given by the TinyLlama model.}
    \label{gpt-prompt}

\end{figure}

The proposed approach uses the dataset introduced in \cite{chen2024graphwiz} for training and evaluation.
This dataset is made of a set of user queries and answers.
Each query describes a graph textually by giving the number of nodes and the list of edges.
After the description, a mathematical question on the graph is asked.
An answer is associated to each query, this is the answer to the mathematical question.
There are nine categories of questions: triangle, cycle, flow, connectivity, bipartite, hamilton, substructure, shortest and topology.
Since some of the tasks also rely on the understanding of edge features, only four of the nine categories are used: cycle, connectivity, bipartite and hamilton.

Given the structure of the original dataset, it is possible to extract one dataset for each training stage.
Given a query, the description of the graph can be used to create the associated graph.
To do so, the edge list is extracted from the query and for each node, the sentence “This is node {node number}” is given as node feature.
Using the graph’s description as $X_a^i$ and sampling uniformly form the set described in Fig. 2 for
$X_q^i$ builds the dataset for the first training stage.
Extracting the mathematical question from each query for $X_q^i$ and using the associated answer for  $X_a^i$ builds the dataset for the second training stage.


\subsection{Experimental Setup and Implementation}

As far as model implementation goes, $f_{\theta}$ and $E_{\epsilon}$ are implemented as TinyLlama \cite{tiny-llama}, $W$ is implemented as a MLP with 2 layers and GELU \cite{hendrycks2023gaussian} as activation functions and $g_{\theta}$ is implemented as a graph encoder as introduced in \cite{graph-llm} adapted to fit the output requirements as described in previous section.

Both training stages are performed on the datasets mentioned above.
The first stage uses 3 epochs with AdamW \cite{adamw} optimizer and a learning rate of $0.001$.
The second stage uses 3 epochs with AdamW \cite{adamw} optimizer and a learning rate of $0.001$.
In the second stage, the LLM part is trained using LoRA \cite{lora} on the query and key attention projections of every attention layer.

For evaluation, GraphLlava is compared with its underlying LLM, that is TinyLlama fine tuned on the same training dataset.
The task is, for both models, zero-shot graph instruction following.
TinyLlama uses the textual graph representation while GraphLlava uses the graph itself.

\subsection{Quantitative Results}

\begin{figure}
    \centering
    \begin{tikzpicture}
        \begin{axis}[
            ybar,
            bar width=20pt,
            width=0.5\textwidth, 
            height=0.5\textwidth,
            ymajorgrids,
            enlarge x limits=0.005, 
            ymin=0,
            ymax=9500,
            ylabel={Number of samples},
            xlabel={Number of tokens on the graph's description},
            xmin=-50,
            xmax=1650,
            xtick={0, 200, 400, 600, 800, 1000, 1200, 1400, 1600},
            x tick label style={rotate=0, anchor=north},
            legend style={at={(0.5,-0.2)}, anchor=north,legend columns=1},
            ]
 
            \addplot+[
                ybar interval,
                pattern=crosshatch,
                pattern color=blue,
                color=white,
                draw=black,
                area legend,
                postaction={pattern=crosshatch, pattern color=blue}
            ] plot coordinates {(0, 9149) (200, 4841) (400, 3155) (600, 2028) (800, 1194) (1000, 806) (1200, 582) (1400, 283) (1600, 0)};

            \addplot+[
                ybar interval,
                pattern=crosshatch,
                color=white,
                pattern color=orange,
                draw=black,
                area legend,
                postaction={pattern=crosshatch, pattern color=orange}
            ] plot coordinates {(0, 6092) (200, 3356) (400, 2208) (600, 1423) (800, 938) (1000, 629) (1200, 465) (1400, 259) (1600, 0)};

            \addplot+[
                ybar interval,
                pattern=north east lines,
                color=white,
                pattern color=chad-green,
                draw=black,
                area legend,
                postaction={pattern=north east lines, pattern color=chad-green}
            ] plot coordinates {(0, 5005) (200, 2753) (400, 1873) (600, 1227) (800, 784) (1000, 530) (1200, 451) (1400, 235) (1600, 0)};

            \legend{Neither, TinyLlama choosen, GraphLlava choosen}
\end{axis}
\end{tikzpicture}
    \caption{Repartition of GPT-4 choices in function of the graph's size.
    The different questions of the test dataset are batched according to the graph's textual description size.
    GPT-4-as-judge metric is reported for each batch.}
    \label{fig:gpt-judge-toks}
\end{figure}

\begin{figure}
    \begin{tikzpicture}[scale=0.9]
    \begin{axis}[
            bar width=20pt,
            width=0.5\textwidth, 
            height=1.2\textwidth,
            ymajorgrids,
            enlarge x limits=0.005, 
            ymin=0,
            ymax=9500,
            ylabel={Number of samples},
            xlabel={Number of tokens on the graph's description},
            xlabel style={yshift=-10},
            xmin=-50,
            xmax=1650,
            xtick={0, 200, 400, 600, 800, 1000, 1200, 1400, 1600},
            x tick label style={rotate=-45, anchor=north west},
            legend style={at={(0.5,-0.1)}, anchor=north,legend columns=1},
            ]
            \addplot[
                ybar interval,
                fill=babyblueeyes,
                draw=black,
                area legend,
            ] plot coordinates {(0, 9149) (200, 4841) (400, 3155) (600, 2028) (800, 1194) (1000, 806) (1200, 582) (1400, 283) (1600, 0)};

            \addplot+[
                color=orange,
            mark options={thick},
            mark size=6pt,
            mark=halfsquare left*,
            clip=false,
            ] coordinates{(100, 6501.0) (300, 3287.0000000000005) (500, 0.0) (700, 0.0) (900, 0.0) (1100, 0.0) (1300, 0.0) (1500, 0.0)};
 
            \addplot+[
            color=chad-green,
            mark options={thick=50},
            mark size=6pt,
            mark=halfsquare right*,
            clip=false,
            ] coordinates {(100, 5258.000000000001) (300, 3160.9999999999995) (500, 2128.0) (700, 1391.0) (900, 771.0) (1100, 502.99999999999994) (1300, 384.0) (1500, 266.0) };
 
            \addplot+[
                color=red,
                ultra thick,
                dotted,
            ] coordinates {(100, 4553) (300, 2423) (500, 1614) (700, 1007) (900, 611) (1100, 397) (1300, 306) (1500, 132)};
 
            \legend{Total Number of samples, Number of samples correctly classified by TinyLlama, Number of samples correctly classified by GraphLlava, Number of samples correctly classified by the random classification}
\end{axis}
\end{tikzpicture}
 
    \caption{Number of correctly answered questions in function of the graph size.
    Similar to Fig. \ref{fig:gpt-judge-toks}, the questions are batched according to the length of the textual descrption of the graphs they are related to.
    The generation task is viewed as a classification task and the accuracy is reported.}
    \label{fig:accuracy}
\end{figure}

In order to measure the quality of the produced answers, GPT-4 is leveraged to judge the results.
For a given question and graph, GPT-4 has to choose between two outputs, the first one is generated by GraphLlava, the second one is generated by TinyLlama.
GPT-4 is prompted with the prompt described in Fig. \ref{gpt-prompt}.
We name this metric GPT-4-as-judge.

Fig. \ref{fig:gpt-judge-toks} shows GPT-4-as-judge metric for different size of the input graph.
The study reveals that, using this metric, for larger graphs, GraphLlava outperforms TinyLlama.
Fig. \ref{fig:gpt-judge-tasks} shows that GraphLlava outperforms TinyLlama in all instructed tasks.
This is also true on average as GPT-4 chooses GraphLlava 56\% of the time, TinyLlama 37.5\% of the time and neither of them 6.5\% of the time.

In order to have another objective metric, we leverage the fact that the dataset is made of questions whose overall answer can be `yes' or `no'.
The generation task is thus perceived as a classification task.
Each answer produced by a model is classified as either a `yes' or a `no' answer and the accuracy of each model is computed.

Fig. \ref{fig:accuracy} shows the accuracy computation for each cluster of graphs comprised in defined graph size range.
The results show that TinyLlama's accuracy plummets for graphs whose description is over 400 tokens while GraphLlava's remains constantly above random.
This due to the fact that GraphLlava uses an embedding representation of the input graph that has a constant size when projected in the textual embedding space and is thus agnostic to changes in graph size.
TinyLlama, on the other hand, relies on the textual description.
When this textual description is too big, the model's context is overwhelmed and it is not anymore able to understand the user's instruction leading to catastrophic accuracy.

On average, GraphLlava scores 62.90\% accuracy, TinyLlama 44.41\% and a randomly generated answer (with 50\% chances of answering `yes') scores 50.11\%.
GraphLlava approach is thus significantly better than TinyLlama on the testing dataset for the accuracy metric.

For both metrics, the GraphLlava approach performs significantly better than the underlying LLM alone, suggesting that the provided graph embeddings allow the model to better understand the graph structure.
GraphLlava is also able to scale to larger graphs, maintaining an above random accuracy where its underlying LLM's collapses for graphs bigger than 400 tokens.

\subsection{Qualitative Results}

Qualitatively speaking, GraphLlava provides a more consistent output in both negative and positive examples for larger graphs.
When larger graphs are inputed such as in Fig. \ref{qualit_neg}, TinyLlama tends to hallucinate and even totally forget about the question as the input context becomes too long. GraphLlava does not have this problem as the encoder compresses every graph in a fixed number of embeddings thus allowing the LLM to perform better.

\begin{figure}
\centering
    \begin{tikzpicture}[scale=0.95]
        \begin{axis}[
        ybar stacked,
    ylabel={Number of Samples},
    ymajorgrids,
    xlabel={Tasks},
    xtick=data,
    ymin=0,
    ymax=7500,
    height=8cm,
    bar width=20pt,
    enlarge x limits=0.2,
    axis line style={black},
    xtick style={black},
    ytick style={black},
    legend style={at={(0.5,-0.2)}, anchor=north,legend columns=1},
    symbolic x coords={cycle, connectivity, bipartite, hamilton},
    tick label style={font=\footnotesize},
    ylabel style={font=\footnotesize},
    xlabel style={font=\footnotesize},
    title style={font=\footnotesize},
    ]
            ]
            \addplot+[
                ybar,
                pattern=north east lines,
                color=white,
                pattern color=chad-green,
                draw=black,
                area legend,
                postaction={pattern=north east lines, pattern color=chad-green}
            ] plot coordinates {(cycle, 4334) (connectivity, 3217) (bipartite, 2537) (hamilton, 2818)};
            \addplot+[
                 ybar,
                pattern=crosshatch,
                pattern color=orange,
                draw=black,
                area legend
            ] plot coordinates {(cycle, 983) (connectivity, 635) (bipartite, 253) (hamilton, 613) };

            \addplot+[
                ybar,
                pattern=crosshatch,
                pattern color=blue,
                draw=black,
                area legend,
            ] plot coordinates {(cycle, 1687) (connectivity, 2214) (bipartite, 2324) (hamilton, 423) };
        \legend{GraphLlava choosen, Tiny-Llama choosen, Neither}
        \end{axis}
    \end{tikzpicture}
    \caption{Repartition of GPT4 choice according to the dataset's tasks.}
    \label{fig:gpt-judge-tasks}
\end{figure}

\begin{figure}
\centering

\begin{tcolorbox} 
\textbf{Textual Representation of the Graph}: The nodes are numbered from 0 to 7, and the edges are: (0, 6) (0, 7) (0, 2) (0, 3) (0, 5) (1, 4) (1, 2) (1, 7) (2, 4) (2, 6) (2, 3) (3, 6) (5, 7) (5, 6) (6, 7). \\
\textbf{Question}: Is there a cycle in this graph? \\
\textbf{Ground Truth}: There is a cycle in this graph. One possible cycle is [0-6-7-0], which can be formed by following the edges (0, 6), (6, 7), and (7, 0). Another possible cycle is [2-4-1-7-2], which can be formed by following the edges (2, 4), (4, 1), (1, 7), and (7, 2). There may be other cycles as well, but these are sufficient to show that the graph is not acyclic.\#\#\# Yes. \\
\textbf{GraphLlava}: \#\#\# Yes, there is a cycle in this graph. \\
\textbf{TinyLlama}: Yes, there is a cycle in the given graph. The graph has a directed edge from node 0 to node 6, a directed edge from node 0 to node 7, and a directed edge from node 1 to node 4, which completes a cycle.

\end{tcolorbox}
\caption{Small context output example. Both models have the right response even though TinyLlama hallucinates the explanation.}
    \label{qualit}

\end{figure}

When the context length is shorter, TinyLlama usually attempts to provide an explanation of the result as shown in Fig \ref{qualit}.
GraphLlava, on the other hand, provides shorter answers this is mainly due to the cropping done at training so that the length of the labels can fit on limited GPU memory.
This transformation often leads to data only containing the most important part of the answer and removing the explanations.

\begin{figure}
\begin{tcolorbox}

\textbf{Textual Representation of the Graph}: The nodes are numbered from 0 to 44, and the edges are: (0, 13) (0, 18) (0, 1) (0, 42) (1, 21) (1, 36) (1, 27) (2, 5) (2, 17) (3, 15) (5, 22) (5, 20) (5, 38) (5, 26) (5, 31) (5, 30) (6, 14) (6, 22) (6, 18) (6, 12) (7, 22) (8, 28) (9, 41) (9, 10) (9, 43) (9, 19) (12, 35) (13, 42) (14, 15) (14, 22) (15, 26) (16, 36) (17, 25) (18, 40) (19, 41) (19, 21) (19, 34) (19, 39) (21, 36) (21, 32) (24, 41) (24, 33) (26, 37) (26, 39) (26, 28) (29, 32) (29, 41) (31, 33) (32, 42) (32, 37) (32, 36) (33, 35) (34, 41) (35, 39) (35, 43) (38, 40) (41, 43).\\
\textbf{Question}: Is there a path between node 28 and node 11? \\
\textbf{Ground Truth}: There is no path between node 28 and node 11. Node 28 is not in the same connected block as node 11.\\
\textbf{GraphLlava}: There is no path between node 28 and node 11.\\
\textbf{TinyLlama}: Yes, the given graph is an undirected graph, and the nodes are numbered from 0 to 44, and the edges are:(0, 13) (0, 18) (0, 1) (0, 42) (1, 21) (1, 36) (1, 27) (2, 5) (2, 17) (3, 15) (5, 22) (5, 20) (5, 38) (5, 26) (5, 31) (5, 30) (6, 14) (6, 22) (6, 18) (6, 12) (7, 22) (8, 28) (9, 41) (9, 10) (9, 43) (9, 19) (12, 35) (13, 42) (14, 15) (14, 22) (15, 26) (16, 36) (17, 25) (18, 40) (19, 41) (19, 21) (19, 34) (19, 39) (21, 36) (21, 32) (24, 41) (24, 33) (26, 37) (26, 39) (26, 28) (29, 32) (29, 41) (31, 33) (32, 42) (32, 37) (32, 36) (33, 35) (34, 41) (35, 39) (35, 43) (38, 40) (41, 43). Is there a path between node 28 and node 11?

\end{tcolorbox}
\caption{Long context output example. TinyLlama completely hallucinates the answer as the input context is too long for for it.}
    \label{qualit_neg}

\end{figure}

\section{Limitations}

\subsection{Architectural limitations}

The graph encoder presents a fundamental limitation in its design.
Although efficient for reducing the context length used by the graph description, it compresses the graph information to a finite, predefined size.
This compression ends up being important if the node features are too long or if the number of nodes grows too much.
The information loss results in hallucination in given answers.
A query-aware encoder could thus be designed to take into account the parts of the graph where the compression can be higher and those that need to keep most of their information.

\subsection{Training data limitations}

As far as the training pipeline is concerned, the approach has a main drawback that is the need of a conversational dataset.
Building a conversation between a user and an assistant can be very expansive and is hard to automate.
The authors in \cite{llava} leverage the textual description of an image enhanced with bounding boxes positions and a GPT-4 to generate such conversations.
When images are replaced by graphs, the dataset generation is confronted to the following limitations:
\begin{itemize}
    \item Obtaining a textual representation that does not end into a lengthy context is a complex task especially if the graph has complex node features such as text.
    \item As graphs get more and more complex, the length of the textual description, even of a well-compressed one, gets bigger. This can lead to queries with context that are too long for LLMs and thus generate incoherent conversations.
    \item LLMs are known for having trouble using textual description of graphs to solve complex tasks \cite{llm-vs-graph}.
\end{itemize}

Generating the conversation using GPT-4 as a teacher is thus risky since the resulting approach could would have GPT-4's graph understanding as an upper bound.
Moreover, scaling the approach to more complex graphs could pose a serious problem if no dataset is available or if available generative models do not perform well on such graphs.

\subsection{Experimental Limitations}

Due to limited GPU computing power, the experiment presented in the previous sections uses a small language model as LLM backbone in the GraphLlava architecture.
Despite giving promising results, scaling the architecture with a larger language model would be essential to have a better understanding of the efficiency of the approach.
Limited GPU resources also lead to a small training and evaluation sequence length (512 tokens), meaning that the model produces only simple answers and is not trained to give too much detail.
From the evaluation perspective, the output is cropped at a certain length thus introducing noisy results as input for GPT-4-as-judge metric.

As far as the computed metrics go, the GPT-4-as-judge metric is a way of evaluating qualitative results in an automated way.
This approach is, of course, limited by the model's understanding of the textual representation of the graph and introduces GPT-4 as an upper bound to the model's performance evaluation.
Although acceptable at this stage, finding a better metric for future works would be an essential step.

\section{Conclusion}

This works investigates the implementation of a multimodal architecture that enhances LLMs with graph understanding for general graph instruction following tasks.
The proposed approach leverages a graph encoder to produce embeddings that are converted to textual embeddings and injected into a LLM architecture.
Combined with the embeddings of the instruction, the new graph representation is used by the LLM to generate a relevant answer.

This approach is implemented using TinyLlama as LLM as well as an already pretrained graph encoder.
It achieves better overall performances in both accuracy and GPT-4-as-judge metric than its underlying LLM conterpart especially for larger graphs whose textual descriptions overwhelm TinyLlama's context.
This approach provides a scalable and robust way of combining graphs and LLM for instruction following tasks.

Although giving promising results, due to limited GPU resources, the model is implemented using a small language model and a quite limited dataset.
For future works, training a GraphLlava architecture with a stronger backbone LLM and with a more complex dataset while improving the encoder architecture is essential.

\printbibliography

\end{document}